# Synchronous phase-demodulation of concentric-rings Placido mires in corneal topography and wavefront aberrometry (theoretical considerations)

**M. Servin***

*Centro de Investigaciones en Optica A. C., Loma del Bosque 115, 37150, Leon Guanajuato, Mexico.*
*[*]mservin@cio.mx*

**Abstract:** This paper presents a digital interferometric method to demodulate Placido fringe patterns. This method uses a computer-stored conic-wavefront as reference carrier. Even though, Placido mires are widely used in corneal topographers. This is not however a paper on corneal topography and/or its clinical use. This paper focuses on the theoretical aspects to phase-demodulate Placido mires using synchronous interferometric techniques. Placido patterns may also be applied to test optical wavefronts using a Placido-Hartmann opaque plate with periodic annular apertures. This test is sensitive to the radial slope of the measuring wavefront. Another wavefront testing approach may use a Placido-Hartmann-Shack screen with a periodic array of toroidal lenslets. This periodic screen is sensitive to the wavefront's radial-slope at the focal plane of the lenslets. In brief, digital interferometric methods are herein applied for the first time to demodulate conic-carrier Placido images. (Patent pending at the USPTO).

13. V. N. Mahajan, "Zernike Polynomials and Wavefront Fitting," in .*Optical Shop Testing*, 3th ed., D. Malacara ed., John Wiley & Sons, Inc. (2007).

**1. Introduction**

Periodic concentric-rings Placido fringe patterns are commonly used to measure the topography irregularities of the human cornea since 1880 [1]. Modern corneal topographers still use Placido's targets, or some variations of it, for example: the ATLAS 9000, the Tomey TMS, the Astra Max, the Magellan Mapper, the Keratron, the Topolyzer, the EyeSys 3000 and the EyeSys Vista [1,2]. The Placido target is placed in front of a human eye and its reflected image is digitized with a camera placed at a center hole of the Placido's mire. The reflected Placido image is phase-modulated by the corneal irregularities. The geometry of the corneal topographer determines the radial-slope sensitivity $s$ of the apparatus [1-7,10]. Then this Placido image is *phase de-modulated* to estimate the radial slope of the corneal topography with respect to its closest sphere [1-7,10]. Papers focusing on Placido phase-demodulation are uncommon [2,7,10]; nevertheless some are disclosed in Patents [3-6]. Standard Placido demodulation locates the fringe edges of the Placido image using intensity-image processing techniques [3-7]. A set of local distances between the Placido image and its real-valued reference are estimated [3-7]. To increase the number of estimated slope-points, more complicated Placido targets have been proposed [3-7]. Some of these newer Placido patterns are more sensitive to the angular corneal deformations [2-7]. In spite of this, the topographic-slope is not known at every pixel of the Placido image [3-7,10]. Finally this sparse set of estimated slope-points is integrated to obtain the topography deformations of the testing cornea [2-7].

Today Placido image demodulation relies on slope estimations at a sparse set of points [2-7,10]. In contrast synchronous interferometric methods provide holographic phase estimation at every pixel of the fringe pattern domain [8,11,12]. Different kinds of fringe patterns need different wavefront references for their synchronous demodulation. For example linear demodulation uses a plane wavefront as reference [8,11], while pixelated interferometry uses a more complicated wavefront carrier [12]. This paper presents a new digital interferometric method for Placido images which uses a conic wavefront as reference. As far as I know, this is the first time that conic interferometry is used to demodulate Placido images.

Corneal topography is not the only application of Placido patterns. One may use a Placido-Hartmann screen to test optical wavefronts. A Placido-Hartmann screen is an opaque screen with annular periodic apertures. Then as in the Hartmann plate test [9], one digitizes at $d$ meters the Placido's shadow. Using the synchronous interferometric method herein described, one may estimate dense radial-slope of the wavefront at every pixel of the Placido plate's shadow.

Another possible application for Placido patterns in wavefront testing is obtained by modifying the Hartmann-Shack experimental set-up [9]. In the Hartmann-Shack test, one uses a square array of lenslets [9]. The testing image of the Hartmann-Shack square lenslets is taken at the focal distance $f$ of the lenslets [9]. One may modify this experimental set-up using a periodic annular (toroidal) lenslet array instead of a lenslets matrix [10]. This periodic annular lenslets pattern would be sensitive to the radial slope of the wavefront [10].

Note that this is not a paper on corneal topography and/or its clinical applications. Neither this paper pretends to compare Placido-based corneal topographers against: Slit-Scanning Topography, Scheimplug Imaging, Ultrasound Digital Topography, or Optical Coherence Topography [1,2]. The sole purpose of this paper is to communicate an accurate and efficient synchronous demodulation method for Placido images in general, using a conic-wavefront as reference. This digital interferometric method may be implemented by convolution filters in the fringe pattern space, or in the Fourier domain. The applications of Placido targets in corneal topography and wavefront testing are mentioned solely for motivation purposes.

## 2. Synchronous demodulation of Placido images using a conic-wavefront reference

Although most concentric rings Placido reference patterns are binary. To simplify our notation we will model them as a continuous cosine function,

$$\text{Placido}(\rho_1) = 1 + \cos\left(\frac{2\pi}{P}\rho_1\right), \quad \rho_1 = \sqrt{x_1^2 + x_2^2}. \tag{1}$$

Where $(x_1, x_2)$ are the coordinates at the Placido source plate. The constant $(2\pi/P)$ is its radial spatial frequency. The Placido screen in Eq. (1) is then phase-modulated using an apparatus sensitive to the slope of he physical variable of interest; either wavefront aberrations or corneal deformations. Then one obtains the following phase-modulated Placido image,

$$I(\rho,\theta) = a(\rho,\theta) + b(\rho,\theta)\cos\left[\frac{2\pi}{P}\rho + s\frac{\partial\varphi(\rho,\theta)}{\partial\rho}\right]. \tag{2}$$

The sensitivity $s$ in Eq. (2) depends on the experiment: 1) In Placido-Hartmann wavefront testing, $s$ is proportional to the distance $d$ from the testing plate to its shadow image [9]; 2) In Placido-Hartmann-Shack wavefront testing, the sensitivity $s$ is related with the focal distance $f$ of the lenslets array [9,10]; 3) Finally $s$ depends on the corneal topographer's geometry [2-7]. Space $(\rho,\theta)$ is at the camera's CCD plane. The background of the fringes is $a(\rho,\theta)$, and its contrast $b(\rho,\theta)$. Finally $\partial\varphi(\rho,\theta)/\partial\rho$ may be either, the corneal radial-slope deformation or the wavefront radial-slope aberration.

To demodulate the Placido image in Eq. (2) by digital interferometry we must first multiply it by a conic-wavefront reference $\exp[-i(2\pi\rho/P)]$, *having the same spatial frequency and center as the Placido image in Eq. (2).*

$$I(\rho,\theta)\exp\left(-i\frac{2\pi}{P}\right) = \left\{a(\rho,\theta) + b(\rho,\theta)\cos\left[\frac{2\pi}{P}\rho + s\frac{\partial\varphi(\rho,\theta)}{\partial\rho}\right]\right\}\exp\left(-i\frac{2\pi}{P}\rho\right). \tag{3}$$

The cosine signal may be decomposed into its two complex exponentials obtaining,

$$I\exp\left[i\frac{2\pi}{P}\rho\right] = \left\{a + \frac{b}{2}\exp\left[i\left(\frac{2\pi}{P}\rho + s\frac{\partial\varphi}{\partial\rho}\right)\right] + \frac{b}{2}\exp\left[-i\left(\frac{2\pi}{P}\rho + s\frac{\partial\varphi}{\partial\rho}\right)\right]\right\}\exp\left(-i\frac{2\pi}{P}\rho\right). \tag{4}$$

Where the polar axis $(\rho,\theta)$ were omitted. Finally this product may be re-written as

$$I\exp\left(-i\frac{2\pi}{P}\rho\right) = a\exp\left(-i\frac{2\pi}{P}\rho\right) + \frac{b}{2}\exp\left[-i\left(2\frac{2\pi}{P}\rho + s\frac{\partial\varphi}{\partial\rho}\right)\right] + \frac{b}{2}\exp\left(is\frac{\partial\varphi}{\partial\rho}\right) \tag{5}$$

This result has three terms: the first term is the conic-wavefront $a(\rho,\theta)\exp[-i(2\pi/P)\rho]$; the second term is a wavefront having twice the reference carrier $2(2\pi/P)\rho$; the third term is the desired analytical-signal $(b/2)\exp[i\,s(\partial\varphi/\partial\rho)]$. Finally we must filter-out the two high-frequency conic wavefronts in Eq. (5) using a low-pass filter (LPF[.]) to obtain,

$$\text{LPF}\left\{I(\rho,\theta)\exp\left(-i\frac{2\pi}{P}\right)\right\} = \frac{b(\rho,\theta)}{2}\exp\left(is\frac{\partial\varphi(\rho,\theta)}{\partial\rho}\right). \tag{6}$$

This equation shows the proposed method to demodulate a Placido image using digital interferometry. Whenever the three spectra in Eq. (5) are well separated we obtain, error-free, the desired baseband analytical signal (Eq. (6)). The condition for spectral separability, and therefore for error-free demodulation, is the following supremum (least upper bound) for the testing radial-slope $\partial \varphi(\rho,\theta)/\partial \rho$,

$$\sup\left[\frac{\partial}{\partial \rho}\left(s\frac{\partial \varphi(\rho,\theta)}{\partial \rho}\right)\right] < \frac{2\pi}{P}. \tag{7}$$

The final step to obtain the desired radial slope $\partial \varphi(\rho,\theta)/\partial \rho$ from Eq. (6) is,

$$\frac{\partial \varphi(\rho,\theta)}{\partial \rho} = \frac{1}{s}\tan^{-1}\left[\frac{\text{Im}\{\text{LPF}[I(\rho,\theta)\exp(-i2\pi\rho/P)]\}}{\text{Re}\{\text{LPF}[I(\rho,\theta)\exp(-i2\pi\rho/P)]\}}\right]. \tag{8}$$

Where Im[.] and Re[.] are operators which take the real and imaginary parts of their argument. Eq. (8) gives in a single formula, the proposed phase-demodulation method. The radial-slope in Eq. (8) is dense, that is we obtain radial-slope estimation at every pixel on the measuring field, not just a cloud of sparse slope estimation points [2-7,10]. The phase in Eq. (8) is wrapped modulo $2\pi$, so before radial integration to obtain $\varphi(\rho,\theta)$, we must unwrap it.

All digital synchronous interferometric methods need a wavefront or complex-valued carrier as reference [8,11,12]. In digital linear interferometry one uses a plane wavefront *i.e.* $\exp[i(2\pi/P)(x+y)]$ as reference [8,11]. In digital pixelated interferometry the reference wavefront is composed by 2x2 wavefront unit-cells tiled over the 2D space [12] as,

$$\text{Pixelated\_reference\_wave}(x,y) = \left[\sum_{n\in\mathbb{Z}}\sum_{m\in\mathbb{Z}}\delta(x-2n, y-2m)\right] ** \exp\left[i\begin{pmatrix} 0 & \pi/2 \\ 3\pi/2 & \pi \end{pmatrix}\right] \tag{9}$$

Finally in this paper one uses a conic-wavefront $\exp\left[i(2\pi/P)\sqrt{x^2+y^2}\right]$ as reference. The fringe patterns generated by these complex-valued carriers have different spectra, as well as different phase-demodulating properties. Trying to demodulate a Placido fringe pattern using a plane wavefront [11] or a pixelated wavefront (Eq. (9)) [12] would completely fail; Placido images must be demodulated using the conic-wavefront reference: $\exp[i(2\pi/P)\rho]$.

### 3. A computer generated example

In this section a computer generated example of interferometric Placido image demodulation using a conic-wavefront is presented in the Fourier domain. This demodulation could have also be performed in the image space using a low-pass convolution filter. But looking the spectra of the Placido images is instructive. Here we are not dealing with any concrete application of Placido mires; this is just a theoretical paper on how to demodulate generic Placido fringe patterns. But as mentioned, this conic reference synchronous method is easily applicable to Placido images in corneal topography and wavefront aberrometry. Wavefront aberrometry would need the use of a Placido-Hartmann plate made of periodic annular apertures. Also wavefront aberrometry can be made with a Placido-Hartmann-Shack plate, made of annular toroidal lenslets as pointed out by Carvalho [10].

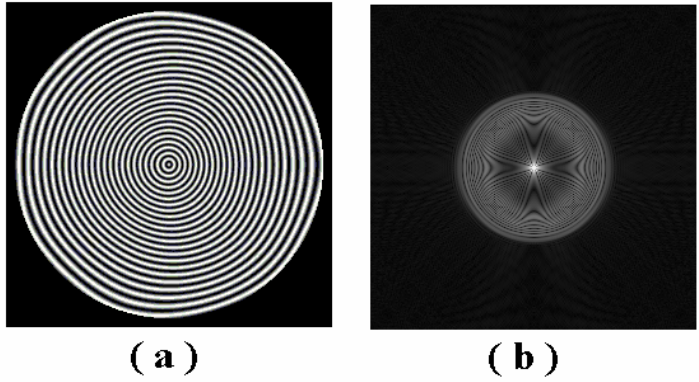

Figure 1. Panel (a) shows the phase-modulated Placido target in Eq. (2). Panel (b) shows the spectrum of the modulated Placido image in panel (a).

Figure 1 shows a modulated Placido image (256x256 pixels) and its frequency spectrum. The spatial frequency of the conic-carrier is $(2\pi/P) = 0.4\pi$ (radians/pixel) along the radial axis $\rho$. Panel 1(b) shows the magnitude of the Fourier spectrum of the Placido image. Please note that both, the positive and negative complex conjugate spectra are superimposed at the donut-shaped region with radius $0.4\pi$ (Eq. (2)). In other words, the two conjugate signals are not spectrally separated as in linear interferometry [11], or as in pixelated interferometry [12]; at first sight, it is not obvious how to separate these two conjugate overlapping spectra.

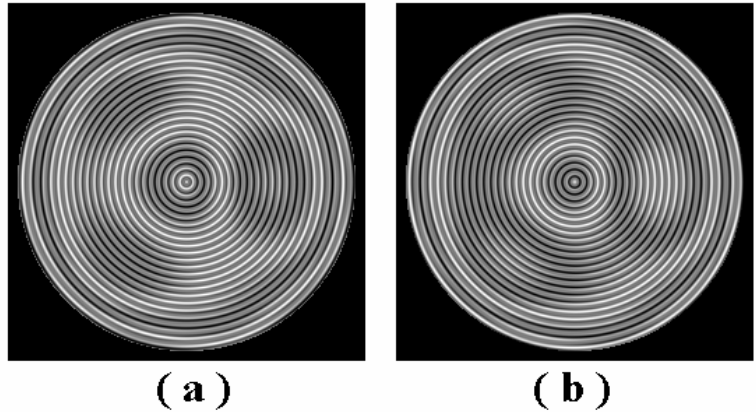

Figure 2. This figure shows the real and the imaginary parts of the product in Eqs. (3-5). Panel (a) shows the real Moiré of the product, while panel (b) the imaginary Moiré.

Figure 2 shows the real and imaginary Moirés obtained by the product of the reference wavefront and the Placido image Eq. (3-5). We may see, as expected, that the real and the imaginary Moirés are in quadrature.

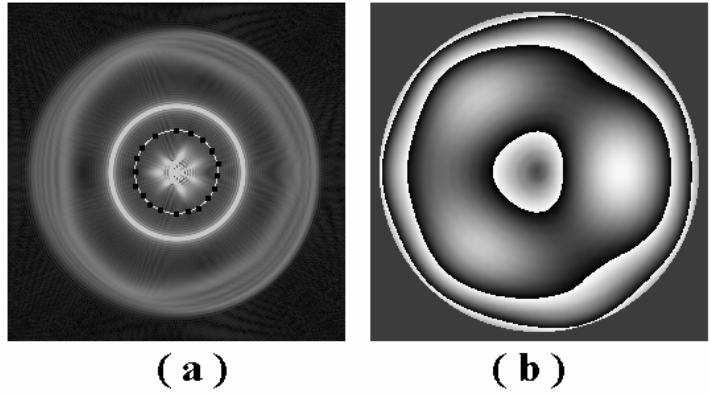

Figure 3. Panel (a) shows the spectrum of the Placido image multiplied by the conic reference wavefront Eq. (3-5). Panel (a) also shows the size of the low-pass-filter as a dotted ring. Panel (b) shows the wrapped phase modulo $2\pi$ given by Eq. (6) and Eq. (8).

Panel 3(a) shows the frequency spectrum of the product between the conic-wavefront reference and the Placido image (Eq. (5)). Note the important fact that: *the two conjugate spectra are now well separated*. The spectrum at the origin is the desired analytical signal: $(b/2)\exp[i\,s(\partial\varphi/\partial\rho)]$. The middle-frequency spectrum corresponds to the conic-wavefront, $a(\rho,\theta)\exp[-i(2\pi\rho/P)]$. Finally the outermost spectral "halo" (the middle term in Eq. (5)) is a modulated conic wavefront having double spatial frequency $2(2\pi\rho/P)$. Panel 3(a) depicts as a dotted circle, the boundary of the LPF[.] used in this simulation (Eq. (6)). The LPF[.] rejects the two high-frequency conic-wavefronts in panel 3(a) (Eq. (5)). Finally panel 3(b) depicts the desired phase-slope $\partial\varphi(\rho,\theta)/\partial\rho$ wrapped modulo $2\pi$ (Eq. (8)).

Please note that an *un-modulated* cosine-profile Placido target (Eq. (1)) do not give a spectral ring-shaped Dirac-delta; only Bessel-profile Placido mires does, because,

$$H\left[1+J_0\left(\frac{2\pi}{P}\rho_1\right)\right] = \delta(\omega_\rho) + \frac{P}{2\pi}\delta\left(\omega_\rho - \frac{2\pi}{P}\right); \quad \rho_1 = \sqrt{x_1^2 + x_2^2},\, \omega_\rho = \sqrt{\omega_x^2 + \omega_y^2}. \quad (10)$$

Where the operator $H(\cdot)$ is the Hankel transform, $\delta(\cdot)$ is the Dirac delta functional, and $\omega_\rho$ is the radial axis in the Fourier domain. This is shown in Fig. 4, where the un-modulated real part of the conic-wavefront reference is shown in panel 4(a) along with its spectrum in panel 4(b). This spectrum resembles a ring Dirac-delta, but as Eq. (10) shows it is not.

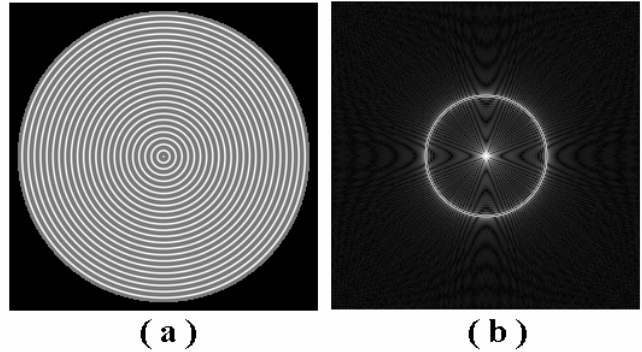

Figure 4. Panel (a) shows the cosine profile un-modulated Placido mire in Eq. (1), and panel (b) shows its spectrum. We can see that the concentric-rings spectrum resembles a ring-shaped Dirac-delta. But as Eq. (10) shows this is not a spectral Dirac-delta ring. To be a ring Dirac-delta the profile of the Placido mire should have a Bessel profile.

Quite often the Placido rings profile is not sinusoidal but binary. In this case we have in general odd and even harmonics of the fundamental concentric-ring signal, as Figure 5 shows. In panel 5(a) we depict a binary non symmetric-profile un-modulated Placido fringe pattern, and in Panel 5(b) its spectrum.

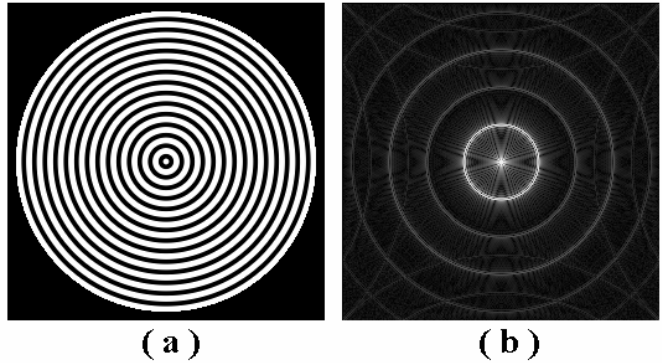

Figure 5. Panel (a) shows a binary-profile un-modulated Placido image and panel (b) shows its spectrum. The spectrum has even and odd harmonics because the binary profile has not a 50% gray-level duty-cycle. We can see that the concentric-rings spectrum now is composed by the fundamental-frequency brightest ring, and its dimmer higher harmonics rings.

Even though we have shown only noise-free Placido images, special mention should be made to the noise in Placido image demodulation. As any other synchronous interferometric method [11,12] the noise rejection of this method is directly proportional to the bandwidth of the desired analytical signal in Eq. (7). If the modulating analytical signal is narrow-band this synchronous method will be highly robust to the corrupting noise of the fringes. This demodulation robustness also applies to the rejection of harmonic signals due to the binary profile normally used in Placido targets. This is shown in Fig. 5 where the harmonic signals of the Placido image have higher frequency content, but they keep far apart from the spectral origin.

**4. Zernike polynomials and the recovery of smooth functions from a radial-slope image**.

Any continuous smooth 2D function within the unit circle may be expanded using orthogonal Zernike polynomials [13], that is

$$\varphi(\rho,\theta) = \sum_{j=1}^{\infty} a_j Z_j(\rho,\theta), \qquad \forall (\rho,\theta) \in [0,1] \times [0,2\pi]. \tag{11}$$

Where the expansion coefficients $a_j$ are given by [13],

$$a_j = \frac{1}{\pi} \int_{\rho=0}^{1} \int_{\theta=0}^{2\pi} \varphi(\rho,\theta) Z_j(\rho,\theta)\, \rho\, d\rho\, d\theta \tag{12}$$

More complicated Placido patterns have been proposed to estimate also angular-slope $\partial \varphi(\rho,\theta)/\partial \theta$ points [2-6]. But as shown here, dense estimation of $\partial \varphi(\rho,\theta)/\partial \rho$ is enough to fully recover any smooth wavefront or topography $\varphi(\rho,\theta)$. That is because all Zernike polynomials are sensitive to the radial derivative [13]; no one disappears by applying the radial operator $\partial/\partial \rho$,

$$\frac{\partial Z_j(\rho,\theta)}{\partial \rho} \neq 0, \quad \forall (j \in \mathbb{N}). \tag{13}$$

In principle, only spatially-dense estimation of the radial-slope $\partial \varphi(\rho,\theta)/\partial \rho$ is needed to obtain $\varphi(\rho,\theta)$ without any knowledge of $\partial \varphi(\rho,\theta)/\partial \theta$. Nevertheless people have searched more elaborated Placido's because standard point-wise phase-demodulation of $\partial \varphi(\rho,\theta)/\partial \rho$ is not dense [2-7]. In this case additional point estimations of $\partial \varphi(\rho,\theta)/\partial \theta$ are useful to obtain a better wavefront or topography $\varphi(\rho,\theta)$ [2-7]. In contrast, interferometric demodulation of Placido images (either in the image or Fourier space) give estimation of the radial slope $\partial \varphi(\rho,\theta)/\partial \rho$ at every pixel within the 2D domain of the fringes, so in principle, there is no need of $\partial \varphi(\rho,\theta)/\partial \theta$.

Keep in mind that, the Placido's image center must coincide with the conic-wavefront reference $\exp[-i(2\pi \rho/P)]$. Otherwise a de-centering erroneous demodulation is obtained.

## 4. Conclusions

A new synchronous digital interferometric method for phase-demodulation of concentric-rings Placido images was presented. This interferometric method uses a conic-wavefront as reference to synchronously demodulate Placido images. This interferometric method estimates the radial-slope at every pixel of the Placido image's domain. Intensity-based phase demodulation methods for Placido, Hartmann or Hartmann-Shack images usually provide only a sparse point-set of slope estimations [2-10]. Additionally intensity-image processing demodulation techniques are more complicated and prone to errors [2-10]. In contrast digital interferometry [8,11,12], provides straightforward holographic demodulation with higher accuracy and at every pixel of the fringe pattern (Eq. (8)). Dense Hartmann plates or dense Hartmann-Shack lenslets plates may be demodulated using linear interferometry [11]. In these cases two slope-fields wavefront aberration along the *x* and *y* axes must be recovered. Using the Placido mire synchronous demodulation herein presented only a single radial slope-field is necessary. Some applications of modulated Placido fringe patterns are: corneal topography, eye wavefront aberrometry, large telescope mirror testing, and adaptive optics. But our main interest in this paper was to present general theoretical aspects of synchronous Placido image demodulation.

## Acknowledgements

I want to thank the financial support of the Mexican National Science Council (CONACYT).